\newread\epsffilein    
\newif\ifepsffileok    
\newif\ifepsfbbfound   
\newif\ifepsfverbose   
\newdimen\epsfxsize    
\newdimen\epsfysize    
\newdimen\epsftsize    
\newdimen\epsfrsize    
\newdimen\epsftmp      
\newdimen\pspoints     
\pspoints=1bp          
\epsfxsize=0pt         
\epsfysize=0pt         
\def\epsfbox#1{\global\def\epsfllx{72}\global\def\epsflly{72}%
   \global\def\epsfurx{540}\global\def\epsfury{720}%
   \def\lbracket{[}\def\testit{#1}\ifx\testit\lbracket
   \let\next=\epsfgetlitbb\else\let\next=\epsfnormal\fi\next{#1}}%
\def\epsfgetlitbb#1#2 #3 #4 #5]#6{\epsfgrab #2 #3 #4 #5 .\\%
   \epsfsetgraph{#6}}%
\def\epsfnormal#1{\epsfgetbb{#1}\epsfsetgraph{#1}}%
\def\epsfgetbb#1{%
%
%
\openin\epsffilein=#1
\ifeof\epsffilein\errmessage{I couldn't open #1, will ignore it}\else
%
%
   {\epsffileoktrue \chardef\other=12
    \def\do##1{\catcode`##1=\other}\dospecials \catcode`\ =10
    \loop
       \read\epsffilein to \epsffileline
       \ifeof\epsffilein\epsffileokfalse\else
%
%
          \expandafter\epsfaux\epsffileline:. \\%
       \fi
   \ifepsffileok\repeat
   \ifepsfbbfound\else
    \ifepsfverbose\message{No bounding box comment in #1; using 
defaults}\fi\fi
   }\closein\epsffilein\fi}%
%
%
\def\epsfclipstring{}
\def\epsfsetgraph#1{%
   \epsfrsize=\epsfury\pspoints
   \advance\epsfrsize by-\epsflly\pspoints   
   \epsftsize=\epsfurx\pspoints
   \advance\epsftsize by-\epsfllx\pspoints
%
%
   \epsfxsize\epsfsize\epsftsize\epsfrsize
   \ifnum\epsfxsize=0 \ifnum\epsfysize=0
      \epsfxsize=\epsftsize \epsfysize=\epsfrsize
      \epsfrsize=0pt  
%
arithmetic!
%
     \else\epsftmp=\epsftsize \divide\epsftmp\epsfrsize
       \epsfxsize=\epsfysize \multiply\epsfxsize\epsftmp
       \multiply\epsftmp\epsfrsize \advance\epsftsize-\epsftmp
       \epsftmp=\epsfysize
       \loop \advance\epsftsize\epsftsize \divide\epsftmp 2
       \ifnum\epsftmp>0
          \ifnum\epsftsize<\epsfrsize\else
             \advance\epsftsize-\epsfrsize \advance\epsfxsize\epsftmp \fi
       \repeat
       \epsfrsize=0pt
     \fi
   \else \ifnum\epsfysize=0
     \epsftmp=\epsfrsize \divide\epsftmp\epsftsize
     \epsfysize=\epsfxsize \multiply\epsfysize\epsftmp
     \multiply\epsftmp\epsftsize \advance\epsfrsize-\epsftmp
     \epsftmp=\epsfxsize
     \loop \advance\epsfrsize\epsfrsize \divide\epsftmp 2
     \ifnum\epsftmp>0 
        \ifnum\epsfrsize<\epsftsize\else
           \advance\epsfrsize-\epsftsize \advance\epsfysize\epsftmp \fi
     \repeat
     \epsfrsize=0pt
    \else
     \epsfrsize=\epsfysize
    \fi
   \fi
%
%
   \ifepsfverbose\message{#1: width=\the\epsfxsize, 
height=\the\epsfysize}\fi
   \epsftmp=10\epsfxsize \divide\epsftmp\pspoints
   \vbox to\epsfysize{\vfil\hbox to\epsfxsize{%
      \ifnum\epsfrsize=0\relax
        \includegraphics{#1}%
      \else
        \epsfrsize=10\epsfysize \divide\epsfrsize\pspoints 
        \includegraphics{#1}%
      \fi
      \hfil}}%
\global\epsfxsize=0pt\global\epsfysize=0pt}%
%
%
{\catcode`\%=12 
\global\let\epsfpercent=
%
%
\long\def\epsfaux#1#2:#3\\{\ifx#1\epsfpercent
   \def\testit{#2}\ifx\testit\epsfbblit
      \epsfgrab #3 . . . \\%
      \epsffileokfalse
      \global\epsfbbfoundtrue
   \fi\else\ifx#1\par\else\epsffileokfalse\fi\fi}%
%
%
\def\epsfempty{}%
\def\epsfgrab #1 #2 #3 #4 #5\\{%
\global\def\epsfllx{#1}\ifx\epsfllx\epsfempty
      \epsfgrab #2 #3 #4 #5 .\\\else
   \global\def\epsflly{#2}%
   \global\def\epsfurx{#3}\global\def\epsfury{#4}\fi}%
%
%
\def\epsfsize#1#2{\epsfxsize}
%
%

\input harvmac.tex
\scrollmode

\ifx\epsfbox\UnDeFiNeD\message{(NO epsf.tex, FIGURES WILL BE
IGNORED)}
\def\Fig.in#1{\vskip2in}
\else\message{(FIGURES WILL BE INCLUDED)}\def\Fig.in#1{#1}\fi
\def\iFig.#1#2#3{\xdef#1{Fig..~\the\Fig.no}  
\goodbreak\topinsert\Fig.in{\centerline{#3}}%
\smallskip\centerline{\vbox{\baselineskip12pt
\advance\hsize by -1truein\noindent{\bf Fig..~\the\Fig.no:} #2}}
\bigskip\endinsert\global\advance\Fig.no by1}

\def \s {\sigma}

\def \ha {\half}
\def \ov {\over}

\def \a {\alpha}

\def \lr { \lref}

\def \dd {\partial }

\def \k {\kappa}

\gdef \jnl#1, #2, #3, 1#4#5#6{ { #1~}{ #2} (1#4#5#6) #3}
\def \np {  Nucl.  Phys. }
\def \pla { Phys. Lett. A }
\def \mpl { Mod. Phys. Lett. }
\def \prl { Phys. Rev. Lett. }
\def \pr  { Phys. Rev. }
\def \cqg { Class. Quant. Grav. }
\def \jmp { J. Math. Phys. }
\def \ap { Ann. Phys. (NY)} 

\def \grg {Gen. Rel. Grav. }


\lr \worm {G. Cl\'ement, \pr D 51 (1995) 6803.}

\lr \tanabe {Y. Tanabe, Progr. Theor. Phys. 57 (1977) 840.}

\lr \ggk {D.V. Gal'tsov, A.A. Garcia and O.V. Kechkin, \cqg 12 (1995) 2887.}

\lr \djh {S. Deser, R. Jackiw and G. 't Hooft, \ap\ 152 (1984) 220.} 

\lr \stat {G. Cl\'ement, \grg 18 (1986) 137.}

\lr \eml {G. Cl\'ement, \cqg 10 (1993) L49.}
 
\lr \ernst {F.J. Ernst, \pr 168 (1968) 1415.}

\lr \nk {G. Neugebauer and D. Kramer, Ann. der Physik (Leipzig) 24 (1969) 62.}

\lr \mazur {P.O. Mazur, Acta Phys. Polon. B14 (1983) 219; 
A. Eris, M. G\"urses, and A. Karasu, \jmp 25 (1984) 1489.}

\lr \exact {D. Kramer, H. Stephani, M. MacCallum, and E. Herlt, {\it Exact
Solutions of the Einstein Field Equations\/}, CUP, 1980.}

\lr \nuph {G. Cl\'ement, \np B 114 (1976) 437.}

\lr \cg {A. Comtet and G. Gibbons, \np B 299 (1988) 719.}

\lr \dema {S. Deser and P.O. Mazur, \cqg 2 (1985) L51.}

\lr \virb {K.S. Virbhadra, preprint gr-qc/9408035 (1994).}

\lr \mcle {R.G. McLenaghan and N. Tariq, \jmp 18 (1975) 2306.}

\lr \tupper {B.O.J. Tupper, \grg 7 (1976) 479.} 

\lr \mcin {C.B.G. McIntosh, \grg 9 (1978) 277.}

\lr \gsa {J.R. Gott, III, J.Z. Simon and M. Alpert, \grg 18 (1985) 1019.}

\lr  \melvin {M.A. Melvin, \cqg 3 (1986) 117.}

\lr \reznik {B. Reznik, \pr D 45 (1992) 2151.}

\lr \kogan {I.I. Kogan, \mpl A 7 (1992) 2341.}

\lr \balfab {R. Balbinot and A. Fabbri, \cqg 13 (1996) 2457.}

\lr \klosch {T. Kl\"{o}sch and T. Strobl, \cqg 13 (1996) 2395.}

\lr \btz {M. Ba\~{n}ados, C. Teitelboim and J. Zanelli, \prl 69 (1992)
1849; M. Ba\~{n}ados, M. Henneaux, C. Teitelboim and J. Zanelli, \pr D 48
(1993) 1506.}

\lr \dj {S. Deser and R. Jackiw, \ap 153 (1984) 405.} 

\lr \mbh {G. Cl\'ement, \pr D 50 (1994) R7119.}

\lr \cold {K.A. Bronnikov, C.P. Constantinidis, R.L. Evangelista and J.C.
Fabris, preprint gr-qc/9710092; K.A. Bronnikov, G. Cl\'ement, C.P. 
Constantinidis and J.C. Fabris, preprint gr-qc/9801050, to be published in
\pla.}

\lr \er {A. Einstein and N. Rosen, \pr 48 (1935) 73.}

\lr \flat {G. Cl\'ement, \jmp 38 (1997) 5807.}

\lr \rev {S. Carlip, \cqg 12 (1995) 2853.}

\lr \bepa {A.A. Belavin and A.M. Polyakov, JETP Lett. 22 (1975) 245.}

\baselineskip8pt
\Title{\vbox
{\baselineskip 6pt
\hbox{GCR-98/04/01}
{\hbox{
   }}} }
{\vbox{\centerline { The gravitating $\s$ model in 2+1 dimensions:}
\vskip .2in \centerline{black hole solutions}
}}
\bigskip\bigskip\bigskip
\vskip -20 true pt

\centerline { G. Cl\'ement\foot{E--mail: gecl@ccr.jussieu.fr}  and A.
Fabbri\foot{Present address: Department of Physics, Stanford University, 
Stanford CA 94305-4060, USA; E--mail: afabbri1@leland.stanford.edu}}
\smallskip \bigskip
\centerline {\it Laboratoire de Gravitation et Cosmologie Relativistes}
\smallskip
\centerline{\it Universit\'e Pierre et Marie Curie, CNRS/ESA 7065}
\smallskip
\centerline {\it Tour 22/12, Bo\^{\i}te 142 -
4, Place Jussieu - 75252 Paris Cedex 05 - France}

\bigskip\bigskip
\bigskip

\baselineskip12pt
\centerline {\bf Abstract}

\bigskip
We derive and discuss black--hole solutions to the gravitating O(3) $\s$
model in (2+1) dimensions. Three different kinds of static black holes are
found. One of these resembles 
the static BTZ black hole, another is
completely free of singularities, and the last type has the same Penrose
diagram as the (3+1)--dimensional Schwarzschild black hole. We also
construct static and dynamical multi--black hole systems.

\bigskip
\baselineskip8pt
\noindent 
\Date {April 1998}

\noblackbox
\baselineskip 16pt plus 2pt minus 2pt

\vfill\eject

\newsec{Introduction}
Models in lower--dimensional gravity are useful as laboratories where
we can study analytically situations which can often be addressed only
numerically in the full four--dimensional setting. A great impetus in the
study of (2+1)--dimensional gravity came from the discovery \btz\ of
black--hole solutions to cosmological gravity with $\Lambda < 0$ \rev.
Less well known is the existence of black--hole solutions to the coupled
Einstein--Maxwell equations with a negative gravitational constant, first
pointed out by Kogan \kogan.

\par
Another model which lends itself to the analytical construction of
stationary solutions is the (2+1)--dimensional gravitating O(3) $\s$
model. The reason is that this model admits stationary multi--soliton
solutions classified by an integer winding number. The flat--space soliton
or vortex solutions originally given by Belavin and Polyakov \bepa\ were
first generalized to self--gravitating soliton solutions by Cl\'ement \nuph,
and independently by Comtet and Gibbons \cg. Wormhole solutions to this
model were discussed in \worm. The aim of the present work is to derive
and discuss black--hole and multi--black hole solutions to the gravitating
$\s$ model.

\par
This model is presented in the next section. We show that the coupled
Einstein--$\s$ field equations in three dimensions may be obtained, for a
special (negative) value of the gravitational constant, by dimensional
reduction from a sector of the stationary Einstein--Maxwell equations in
(3+1) dimensions. We then show that a subset of solutions to the
(2+1)--dimensional Einstein--$\s$ theory, depending on a single real
potential, are in one--to--one correspondence with solutions to the
(2+1)--dimensional Einstein--Maxwell theory.
 
\par
In Sect.\ 3 we discuss static and stationary solutions to the
(2+1)--dimensional Einstein-$\s$ theory. A first set of static solutions
are the multi--soliton solutions of \nuph,\cg,\worm. We are concerned in
the present paper with a second set of solutions, which we construct
explicitly in the case of rotationally symmetric fields depending on a
single real potential. We also discuss briefly the extension of these
static solutions to stationary solutions of the (2+1)--dimensional
Einstein--$\s$ equations and, in an Appendix, their
extension to stationary solutions of the (3+1)--dimensional
Einstein--Maxwell equations. 

\par
Sect.\ 4 is devoted to the study of the causal structure of the two
classes of static circularly symmetric solutions constructed in Sect.\ 3.
The solutions of the first class generically have a non--analytical
singularity. However we show that for negative values of the gravitational
constant these solutions may be analytically extended, for an infinite set
of discrete values of an integration constant, to black--hole spacetimes
falling in two subclasses. The black holes of the first subclass have a
Penrose diagram similar to that of a static BTZ black hole, with a
spacelike singularity hidden behind the horizon. The spacetimes of the second
subclass are completely regular, with a Penrose diagram similar to that of
the extreme BTZ black hole. Finally, the solutions of the second class may
also be extended to black holes for negative values of the gravitational
constant, with a Penrose diagram similar to that of the (3+1)--dimensional
Schwarzschild spacetime.

\par
The extension of these rotationally symmetric solutions to multicenter
solutions is discussed in Sect.\ 5. These multi--black hole solutions
generically admit conical singularities. The requirement that the conical
singularities follow geodesics leads to two possibilities. The first is
that the solution is time--independent, with the conical singularities
lying on the horizon(s). The second possibility is that of an explicitly
time--dependent solution describing a dynamical system of freely falling
black holes and conical singularities. We discuss briefly the dynamical
evolution of such two-black hole systems for the three different black
hole species described in Sect.\ 4.  

\newsec{The three--dimensional gravitating $O(3)$ non--linear $\s$ model
and its one--potential sector}

The three--dimensional $O(3)$ non--linear $\s$ model coupled
to gravity is defined by the action
\eqn\sigra{
S={1\ov 2}\int d^3 x \sqrt{|g|}[-{1\ov {\k}}g^{\mu\nu}R_{\mu\nu}
+ g^{\mu\nu}\dd_{\mu}{\vec\phi}\dd_{\nu}{\vec\phi} + \lambda ({\vec\phi}^2 
-\nu^2)], 
}
where $\k=8\pi G_3$, and
the Lagrange multiplier $\lambda$ constrains the isovector field
${\vec\phi}$ to vary on the two-sphere ${\vec\phi}^2=\nu^2$.
Following \worm\ we construct the stereographic map
\eqn\stere{
\phi_1 + i \phi_2 = {{2\nu\psi}\ov{1+ |\psi^2|}}, \ \ \
\phi_3= \nu {{1-|\psi|^2}\ov{1+|\psi|^2}},
}
that projects the sphere ${\vec\phi}^2=\nu^2$ on the complex $\psi$
plane. The resulting field equations are
\eqn\due{
\nabla^2 \psi = {{2\psi^* (\nabla\psi)^2}\ov{1+|\psi|^2}},
}
\eqn\uno{
R_{\mu\nu}= 2\k \nu^2 {{( \dd_{\mu}\psi^* \dd_{\nu}\psi +
\dd_{\nu} \psi^* \dd_{\mu} \psi)}\ov{(1+|\psi|^2)^2}}.
}

\par
As we now show, these equations are in close correspondence 
with the stationary
Einstein--Maxwell equations in (3+1) dimensions. Under the assumption of a 
timelike Killing vector field $\partial_t$, the four--metric and the
electromagnetic field may be parametrized by
\eqn\statu{
ds_{(4)}^2 = f(dt - \omega_m dx^m)^2 - f^{-1}\gamma_{mn}dx^m dx^n
}
\eqn\statd{
F_{m0}^{(4)} = \partial_m v, \qquad F_{(4)}^{mn} = f\gamma^{-1/2}
\epsilon^{mnp} \partial_p u,
}
where the various fields depend only on the spatial coordinates $x^m$ ($m
= 1,2,3$). The complex Ernst potentials are defined as usual by \exact,\ernst\
\eqn\erpot{
{\cal E} = f- |\Phi|^2 + i \chi , \qquad \Phi = v + iu,
}
where
\eqn\twist{
\partial_m\chi = -f^2\gamma^{-1/2}\gamma_{mn}\epsilon^{npq}\partial_p
\omega_q + 2(u\partial_m v - v\partial_m u).
}
The stationary four--dimensional Einstein--Maxwell equations then reduce to the
three--dimensional Ernst equations \ernst\
\eqn\erun{
({\rm Re}\,{\cal E} + |\Phi|^2) \nabla^2 {\cal E} = (\nabla{\cal E}
+ 2\Phi^*\nabla\Phi)\nabla{\cal E},
}
\eqn\erdu{
({\rm Re}\,{\cal E} + |\Phi|^2) \nabla^2 \Phi = (\nabla {\cal E} + 
2\Phi^*\nabla\Phi)\nabla \Phi,
}
\eqn\ertre{
f^2R_{mn}(\gamma) = {\rm Re} 
\left[ {\ha}{\cal E},_{(m}{\cal E}^*,_{n)} 
+ 2\Phi{\cal E},_{(m}\Phi^*,_{n)}
-2{\cal E}\Phi,_{(m}\Phi^*,_{n)} \right], 
}
where the indices $m,n$, as well as $\nabla$ and $\nabla^2$,
refer to the three metric $\gamma_{mn}$.
These equations have been shown to be those of an SU(2,1) $\s$ model
coupled to three--dimensional gravity \mazur. The particular solution  
$\Phi = 0$ of Eq. \erdu\ (stationary vacuum Einstein equations) reduces
the system (2.9)--(2.11) to the Euclidean field equations for an SU(1,1) 
$\s$ model coupled to three--dimensional gravity \exact. Similarly,
the particular solution \tanabe\
\eqn\cova{
{\cal E}=p^2
}
($p$ real constant) of Eq. \erun\ reduces the system \erun\--\ertre\ to the 
Euclidean SU(2) $\sim$ O(3) $\s$--model field equations \due\ and \uno\ 
provided we make the identifications 
\eqn\iden{
\Phi = p\psi, \qquad \gamma_{mn} = g_{mn}, \qquad \k\nu^2 = -1/2.
}
The three--dimensional gravitational constant $G_3$ is then negative,
which is perfectly legitimate: because three--dimensional gravity is
dynamically trivial, the sign of $G_3$ is not fixed a priori \djh. 

\par
Let us also recall that the equations of the stationary Kaluza--Klein
theory, i.e. five--dimensional general
relativity with two Killing vectors, one timelike ($\partial_t$) and one
spacelike ($\partial_5$) reduce, for a special ansatz, to the
three--dimensional O(3) $\s$--model field equations \due\ and \uno\ for
$\k\nu^2 = -2$ \stat. One may wonder whether such a reduction also exists
in the case of Einstein--Maxwell--dilaton theory
\eqn\emd{
S={1\ov 16\pi}\int d^4 x \sqrt{|g_4|}[-R + 2 \dd^{\mu}{\phi}\dd_{\mu}{\phi} 
- {\rm e}^{-2\a \phi} F^{\mu\nu}F_{\mu\nu}]
}
($\phi$ is the dilaton field with coupling constant $\a$) which
interpolates between the Einstein--Maxwell theory (for $\a = 0$) and the
Kaluza--Klein theory (for $\a = \sqrt{3}$). However inspection of the
five--dimensional Killing algebra of the space of stationary solutions to
Einstein--Maxwell--dilaton theory for $\a \neq 0, \sqrt{3}$ \ggk\ shows
that it does not admit an O(3) subalgebra.

\par
A simple class of solutions to the three--dimensional sigma model \sigra\ 
are those depending on a single real potential $\s$. Then general
arguments \nk\ show that this potential can always be chosen to be harmonic,  
\eqn\armo{
\nabla^2 \s = 0,
}
and that the fields ${\vec\phi}$ or $\psi$ follow geodesics in target
space, i.e. large circles on the sphere ${\vec\phi}^2 = \nu^2$
parametrized by
the angle $\s$. Two examples of such large circles are the meridians
\eqn\meri{
{\vec\phi}=(\nu \cos\alpha \sin\s, \nu \sin\alpha \sin\s, 
\nu \cos\s), \qquad \psi = {\rm e}^{i\alpha}\tan{{\s} \ov {2}}
}
($\alpha$ constant) and the equator
\eqn\equa{
{\vec\phi}=(\nu\cos\s, \nu\sin\s, 0), \qquad \psi = {\rm e}^{i\s}.
}
For all these large circles the Einstein equations
\uno\ reduce to
\eqn\scal{
R_{\mu\nu} = \k\nu^2 \dd_{\mu} \s \dd_{\nu} \s.
} 

\par
Equations \armo\ and \scal\ are the field equations for a massless scalar
field coupled to three--dimensional gravity or, equivalently, for
Einstein--Maxwell theory in three dimensions. Indeed, the second group of
Maxwell equations ${F^{\mu\nu}}_{;\nu} = 0$  implies that the dual
$B_{\rho} \equiv \sqrt{|g|}\epsilon_{\mu\nu\rho}F^{\mu\nu}$ is a
gradient, $B_{\rho} = \nu\dd_{\rho}\s$, i.e.
\eqn\elma{
F^{\mu\nu}={{{\nu}}\ov{\sqrt{|g|}}}\epsilon^{\mu\nu\lambda}
\dd_{\lambda} \s.
}
The first group of Maxwell equations then gives the harmonicity condition
\armo, while the Einstein equations for the electromagnetic field give
the Einstein--scalar equations \scal, owing to the identity between the 
energy momentum tensors
\eqn\enmo{
T_{\mu\nu}=-F_{\mu\rho}F_{\nu}^{\rho}+ {1\ov 4}g_{\mu\nu}F_{\rho\lambda}
F^{\rho\lambda}=\nu^2 [\s_{,\mu}\s_{,\nu}-{1\ov 2}g_{\mu\nu}
\s_{,\rho}\s^{,\rho}].
}
It follows that all the known solutions of the Einstein--Maxwell equations
in (2 + 1) dimensions \dema \gsa \melvin \kogan \eml\ lead to solutions of 
the Einstein--$\sigma$ equations \due\ and \uno\ (however the interpretation 
may be somewhat different), and so also (for $\kappa \nu^2 = - 1/2$) to 
solutions of the (3+1)--dimensional
Einstein--Maxwell equations with ${\cal E} = p^2$, i.e. $f =
p^2(1+|\psi|^2)$, $\chi = 0$. In the case of the ``meridian'' ansatz
\meri, the resulting $p = 1$ four--dimensional metric
\eqn\meriq{
ds_{(4)}^2 = {1\ov{\cos^2(\s/2)}}dt^2 - \cos^2(\s/2) \gamma_{mn} dx^mdx^n 
}
is singular for $\s = \pi$ (mod. $2\pi$); the spatial sections of these
spacetimes are thus generically compact with two symmetrical singularities
$\s = \pm \pi$. The other possible large circle ansatz\"{e} lead to
non--static solutions. In the case of the ``equator'' ansatz \equa\ with
$p = 1/\sqrt{2}$, we obtain
\eqn\equaq{
ds_{(4)}^2 = (dt - \omega_m dx^m)^2 - \gamma_{mn}dx^mdx^n
}
where, from Eq.\twist, the (3+1)--dimensional gravimagnetic field 
\eqn\twistm{
\dd_m\omega_n - \dd_n\omega_m = \nu^{-1} F_{mn}
}
is proportional to the (2+1)--dimensional electromagnetic field \elma.

\newsec{Static and stationary circularly symmetric solutions}

The line element of a static (2+1)--dimensional spacetime may always be
parametrized in the form
\eqn\stato{
ds^2= h^2 dt^2 - {\rm e}^{2u} (dx^2 + dy^2).
}   
where the fields $h$, $u$ are time--independent. We also assume the
complex scalar field $\psi$ to be time--independent (the possibility of a
time--dependent $\psi$ shall be investigated at the end of this section).
Then, choosing complex spatial coordinates $\zeta$, $ \zeta^*$, with $\zeta 
\equiv x+iy$, the Einstein equations \uno\ read \worm\
\eqn\compa{
{{\dd^2 h}\ov{\dd\zeta\dd\zeta^*}} =0,
}
\eqn\umpa{
{{\dd^2 u}\ov{\dd\zeta\dd\zeta^*}}=-{{\k\nu^2}\ov{(1+ |\psi|^2)^2}}
\left( |{{\dd\psi}\ov{\dd\zeta}}|^2 + |{{\dd\psi}\ov{\dd\zeta^*}}|^2 \right),
}
\eqn\erpa{
{{\dd}\ov{\dd\zeta}}({\rm e}^{-2u}{{\dd h}\ov{\dd\zeta}})=
-{{4\k\nu^2 h {\rm e}^{-2u}}\ov{(1+ |\psi|^2)^2}} {{\dd\psi^*}\ov{\dd\zeta}}
{{\dd\psi}\ov{\dd\zeta}}.
}

\par
The general solution to Eq.\ \compa\ is
\eqn\hh{
h = {\rm Re}\,w(\zeta)
}
for some analytical function $w(\zeta)$. The case where the function
$w(\zeta)$ is constant, i.e. 
\eqn\hun{
h = 1,
}
was previously treated in \nuph, 
\worm, and independently in \cg. 
In this case Eq.\ \erpa\ shows that $\psi$ is an analytic (or antianalytic)
function of $\zeta$, which also solves Eq.\ \due. The integration of Eq.\
\umpa\ then leads to two classes of multi--soliton solutions, such that the
map $\psi(\zeta)$ covers an integer number of times the sphere $\vec\phi^2
= \nu^2$. The solutions of the first
class are everywhere regular and asymptotic to the multiconical solutions
of vacuum three--dimensional gravity \nuph, the conical singularities of
the vacuum metric ($\delta$--function sources) being smoothed out by the
extended $\s$--model sources; the corresponding solutions of the
(3+1)--dimensional Einstein equations with ${\cal E} = 1$ are $pp$--waves
(see the Appendix). The solutions of the second class are
also regular for $\k < 0$, but now they have two asymptotically conical
regions at spatial infinity connected by $n$ wormholes \worm. For $\k =
0$, both classes of solutions reduce to the (2+1)-dimensional static 
multiconical spacetime. 

\par
We are interested in this paper in the case where $w(\zeta)$ is not
constant. The zeroes of this function ---Killing horizons--- will lead to 
event horizons of the metric
\stato\ if  the spatial metric is regular there. 
 In particular the
circularly symmetric solution must be such that the functions $h$ and $u$
depend only on the radial coordinate, which we may choose to be $x$ ($y$
is then the angular coordinate). Then the harmonic function $h$ is
\eqn\hx{
h = x
}
($w = \zeta$; we have absorbed an arbitrary multiplicative constant in a
time rescaling). From this special choice, the general static solution
with non--constant $w(\zeta)$ may be recovered by a conformal
transformation, see Sect.\ 5. Carrying out on the static
(2+1)--dimensional metric \stato\ with $h = x$ the (locally trivial)
coordinate relabellings $x \rightarrow \rho$, $y \rightarrow z$ and the
Wick rotation $t \rightarrow i\varphi$, and inserting the result in
\statu, we obtain (up to a gauge transformation) the (3+1)--dimensional
metric
\eqn\staxi{
ds_{(4)}^2 = f(\,dt - \omega_3\,d\varphi)^2 - f^{-1}({\rm
e}^{2u}(\,d\rho^2 + dz^2) + \rho^2\,d\varphi^2)
}
(with $f$ and $\omega_3$ given by \erpot\ and \twist\ for ${\cal E} =
p^2$), which is the Weyl form of the stationary axisymmetric metric if
$\varphi$ is an angle. Thus, the static solutions of the three-dimensional
Einstein--$\sigma$ equations with $\k\nu^2 = -1/2$ are in correspondence
with stationary axisymmetric solutions of the four--dimensional
Einstein--Maxwell equations with ${\cal E} = p^2$.

To solve the remaining (2+1)--dimensional Einstein equations, we further 
assume that
the $\s$--model field depends on a single real potential $\s$, so
that Eqs.\ \due\ and \erpa\ reduce to 
\eqn\sig{
h^{-1}\dd_i(h\dd_i\s) = 0,
} 
\eqn\inde{
\dd_{\zeta}({\rm e}^{-2u}\dd_{\zeta}h)= -\k\nu^2 h {\rm e}^{-2u}
(\dd_{\zeta}\s)^2
}
($i = 1,2$). Because $h$ and $u$ depend only on $x$, the left--hand side of 
Eq.\ \inde\ is real. The reality of the right--hand side then implies
\eqn\solsig{
\dd_x\s \dd_y\s = 0,
}
which has only two independent solutions, further restricted by Eq.\ \sig. 

\par
The first solution $\s = \s(x)$ yields
\eqn\foge{
\s=a\ln x.
}
This massless scalar field is generated by a $\delta$--function source,
so that the equivalence \elma\ with three--dimensional Einstein--Maxwell
theory breaks down, as the integrability condition \armo\ is not
identically satisfied. The corresponding $\s$--model field winds indefinitely 
around a large circle of the
two--sphere $\vec\phi^2 = \nu^2$. The resulting solution to Eq.\ \inde\ 
\eqn\sobi{
u={{\k \nu^2 a^2}\ov 2}\ln x + \ln b
}
leads to the spacetime metric
\eqn\spame{
ds^2=x^2dt^2 - b^2 x^{\k \nu^2 a^2} (dx^2 + dy^2)
}
(previously given, in a different parametrization, in \virb).
The only non--vanishing mixed component of the Ricci tensor is, from Eq.\ 
\scal,
\eqn\ric{
R^1_1 = -\k\nu^2a^2b^{-2} x^{-\k\nu^2a^2 - 2}
}
so that there is generically a naked curvature singularity. The Killing 
horizon $x=0$ is at infinite geodesic distance for $\k\nu^2a^2 \le -4$
while, owing to the non--analytical behavior near $x = 0$,
geodesics generically terminate there for $\k\nu^2a^2 > -4$. However, as
we shall show in the next section, the spacetime \spame\ presents regular
horizons for an infinite discrete set of values of the integration
constant $a$.

\par
Now we turn to the second solution of Eq.\ \solsig, $\s = \s(y)$. From Eq.\
\sig\ this results in
\eqn\map{
\s = ny .
}
Remembering that $y$ is an angular coordinate, we see that the $\s$--model
field $\psi(\zeta)$ is single--valued if $n$ is integer. Integration of
Eq.\ \inde\ then gives  
\eqn\usfe{
u= - {{\k n^2\nu^2}\ov 4} x^2 + \ln b,
}
leading to the spacetime metric
\eqn\bh{
ds^2=x^2dt^2 - b^2{\rm e}^{-\k n^2\nu^2 x^2 /2} (dx^2 + dy^2).
}
The associated solution of three--dimensional Einstein--Maxwell theory,
with a radial electrostatic field $F_{01} = -bn\nu x$ corresponding to the
electric charge
\eqn\char{
Q = {1 \ov 2} \oint \sqrt{|g|} F^{\mu\nu} \epsilon_{\mu\nu\lambda}
dx^{\lambda} = \nu \oint d\s = 2\pi n \nu,
}
was previously given in \dema\ \gsa\ \melvin\ \reznik\ \kogan\ \eml. As we 
shall recall in the next section, for $\k < 0$ the spacetime \bh\ is a black
hole with a Penrose diagram similar to that of the Schwarzschild solution
\kogan. Let us also note that when $\kappa$ goes to zero both metrics
\spame\ and \bh\ go over into the rotationally symmetric Rindler metric
\eqn\rin{
ds^2 = x^2 dt^2 - dx^2 - dy^2 .
}

\par
We shall not attempt here a full investigation of the stationary problem
associated with the action \sigra. As in the (2+1)--dimensional 
Einstein--Maxwell case
\dema, a subset of rotating solutions may be generated from the static
circularly symmetric solutions given above by the local coordinate 
transformation
\eqn\rot{
t \rightarrow t - \omega y
}
($\omega$ constant). While the corresponding local transformation on the
(3+1)--dimensional stationary axisymmetric metric \staxi\ $\varphi
\rightarrow \varphi - \gamma z$ is innocuous, the transformation \rot\ may
lead to the appearance of closed timelike curves, because of the
periodicity of $y$. The resulting stationary solutions are discussed in the 
next section.

\par
Other stationary solutions to the Einstein--$\sigma$ model may be obtained
by relaxing the assumption that the complex field $\psi$ is
time--independent to allow for fields $\psi$ depending periodically on time.
We again assume that $\psi$ depends on a single real potential $\sigma$
and that the spacetime metric is static. Then, the ($0,i$) component of
Eq.\ \scal\ gives
\eqn\time{
\dd_t\s \dd_i\s = 0,
}
so that if $\s$ is time--dependent then it must be space--independent and,
from Eq.\ \armo, linear in time,
\eqn\timu{
\s = ct,
}
which indeed corresponds to a periodic $\s$--model field. Note that the
field \timu\ is obtained from \map\ by the interchange $y \leftrightarrow
t$. It follows that the same interchange, accompanied by the continuation
$x^2 \rightarrow -x^2$ and $b^2 \rightarrow -b^2$ so that the new metric
has the correct signature, leads to the static circularly symmetric metric
generated by \timu\
\eqn\magn{
ds^2 = b^2 {\rm e}^{\kappa c^2 \nu^2 x^2/2} (dt^2 - dx^2) -x^2 dy^2 .
}
The associated ``magnetic'' solution of three--dimensional
Einstein--Maxwell theory, with a radial magnetic  field $F_{12} = bc\nu x$,
was first given by Melvin \melvin\ (see also \kogan\ \eml). The metric
\magn\ is regular for $\kappa > 0$ if $b = 1$, and singular, with compact
spatial sections, for $\kappa < 0$. Other stationary solutions (previously
given in the Einstein--Maxwell case in \eml) may be generated from \timu\
\magn\ by the local coordinate transformation \rot, which leads to
single--valued $\s$--model fields only for the discrete values $\omega_n =
n/c$.

\newsec{Analysis of the global causal structure of these solutions} 
In this section we study the causal structure of the static spacetimes
\spame\ and \bh, as well as of their stationary extensions by the local
transformation \rot. We first consider the spacetime metric \spame,
which can be transformed to the conformal gauge metric
\eqn\spama{
ds^2 = \left( {{|\alpha|}\over{b}}r \right)^{2/\alpha} (dt^2 - dr^2) - \alpha^2
r^2  dy^2, 
}
by the coordinate transformation $r = (b/|\alpha|)x^{\alpha}$,
where we have put $\alpha \equiv \kappa \nu^2 a^2/2 \neq 0$. The resulting
Penrose diagram is a triangle bounded by the spacelike side $r = 0$ and
the two lightlike sides $r \pm t = \infty$. To further elucidate the conformal
structure of this family of spacetimes, we transform for $\alpha \neq -2$
to the new radial coordinate $\rho = bx^{\alpha+2}/|\alpha+2|$, leading to the 
Schwarzschild--like form of this solution
\eqn\spams{
ds^2 = f dt^2 - {1 \over f} d\rho^2 -b^2 f^{\alpha} dy^2,
}
with 
\eqn\f{
f(\rho) = \left( {{|\alpha+2|} \over b}\,\rho \right)^{2/(\alpha+2)}
}
(the case $\alpha = -2$, i.e.\ $\k\nu^2a^2 = -4$, shall be considered below, 
Eq. (4.10)).

\par
The metric \spams\ is generically singular for $\rho = 0$, that is $r = 0$ for 
$\alpha < -2$ or $\alpha > 0$, and $r = \infty$ for $-2 < \alpha < 0$; the
Penrose diagrams for the three cases $\alpha < 2$, $-2 < \alpha < 0$,  and
$\alpha > 0$ are shown in Figs.\ 1, 2 and 4. 
The Killing horizon 
corresponds to $\rho  = 0$ for $\alpha > -2$, and to $\rho = \infty$ ($r =
\infty$) for $\alpha < -2$, in which case it is at infinite geodesic distance.
So the Killing horizon is lightlike and at finite geodesic distance only
for $-2 < \alpha < 0$. For $-1 < \alpha < 0$, the curvature scalar \ric\
diverges on this horizon. For $-2 < \alpha < -1$, the curvature
scalar vanishes on the horizon; nevertheless, geodesics generically cannot be 
extended beyond it because of the non--analytical behaviour of the function
$f(\rho)$. However, for 
\eqn\ext{
\alpha = {{2(1-n)} \over n}
}
($n$ integer), $f(\rho) \propto \rho^n$ is analytical and the
spacetime can be extended. To check this we study the geodesic equation
which can be integrated, using the two constants of the motion (energy and
angular momentum)
\eqn\constm{
f\,\dot{t} = E, \qquad b^2f^{\alpha}\,\dot{y} = L,
}    
to
\eqn\geospa{
\dot{\rho}^2 -E^2 + {L^2 \over b^2}f^{1-\alpha} + \varepsilon f = 0,
}
where a dot means derivative with respect to an affine parameter, and 
$\varepsilon = +1$, $0$ or $-1$ for timelike, lightlike or spacelike geodesics.
For the discrete set of values \ext\ of the constant $\alpha$, both $f$
and $f^{1 - \alpha} \propto \rho^{3n-2}$ are analytical so that the
geodesics can be extended beyond $\rho = 0$ by changing $\rho$ to $-\rho$. 

\par
The case $n = 1$ ($\alpha =0$) 
corresponds to the rotationally symmetric Rindler metric \rin\ which, as
is well known, is regular and admits the cylindrical Minkowski spacetime (with
cylindrical spatial sections) as its maximal extension (Fig.\ 3). All other 
values of $n  > 1$ correspond to degenerate horizons. For $n$ odd, 
$n = 3, 5, \cdots$, the Killing field $\dd_t$ becomes spacelike in the
sector II ($\rho < 0$), where the geodesic equation becomes
\eqn\geospas{
\dot{\rho}^2 -E^2 - {L^2 \over b^2}|f|^{1-\alpha} - \varepsilon |f| = 0,
}
showing that geodesics terminate at the spacelike point singularity $\rho
\rightarrow -\infty$ ($|f| \rightarrow \infty$). The Penrose diagram of the
resulting maximal extension, shown in Fig.\ 5, is similar to that of the
static BTZ black hole \btz\ (except that the spacelike singularity of the
BTZ black hole is not a curvature singularity, but a singularity in the
causal structure). An important difference is that, in the
present case, the circle at spacelike infinity ($\rho \rightarrow +\infty$)
is actually a point, the lengths of concentric circles around this
increasing with decreasing ``radius'' $\rho$ as $\rho^{1-n}$, so that the
length of the event horizon $\rho = 0$ is infinite. The vanishing of the
surface gravity at this horizon also implies that the
associated Hawking temperature is zero (such ``cold black holes'', obeying
a similar quantization property, have
also been found as solutions 
to scalar--tensor theories \cold). From Eq.\ \geospa,
almost all geodesics (all except spacelike geodesics with $E = 0$) cross
this horizon to fall towards the singularity $\rho \rightarrow -\infty$.

\par
For $n$ even, $n = 2, 4, \cdots$, $\rho = 0$ is a horizon of even order
connecting two isometrical sectors I ($\rho > 0$) and I' ($\rho < 0$) where the
Killing field $\dd_t$ is timelike. As before, this horizon has infinite
proper length and is crossed by almost all geodesics. The maximal
extension is a geodesically complete infinite sequence of sectors I
and I', leading to a Penrose diagram  (Fig.\ 6), which is
similar to the Penrose diagram for the extreme BTZ black hole $J = Ml$
\btz\ and its $M \rightarrow 0$ limit, the BTZ ``vacuum'' solution
\eqn\vac{
ds^2 = {{r^2} \over {l^2}} dt^2 - {{l^2} \over {r^2}} dr^2 - r^2 d\varphi^2.
}  
Again, we must keep in mind that in our case the circles at infinity
$\rho \rightarrow \pm \infty$ are contracted to points. The similarity
---and the difference--- with the BTZ vacuum solution \vac\ is most
obvious in the special case $n = 2$ ($\alpha = -1$, i.e.\ $\k\nu^2a^2 =
-2$) where $\rho = bx$ and our metric \spame\ takes the form 
\eqn\reg{
ds^2 = x^2 dt^2 - b^2 x^{-2} dx^2 -b^2  x^{-2} dy^2.
}

\par
In the limit $n \rightarrow \infty$, Eq.\ \ext\ goes over into $\alpha =
-2$ ($\k\nu^2a^2 = -4$). The Schwarzschild form of the metric \spame\ is
in this case \spams\ with 
\eqn\fexp{
f(\rho) = {\rm e}^{-2\rho/b}
}
($\rho = -b \ln x$). The Penrose diagram is the same as for $\alpha < -2$
(Fig.\ 1). The Killing horizon $\rho \rightarrow +\infty$ is at infinite
geodesic distance, while only spacelike geodesics ($\varepsilon = -1$) reach
the singularity $\rho \rightarrow - \infty$, massive or massless test
particles being repelled by an exponentially rising potential barrier. 

\par
Performing on \spame\ the local coordinate transformation \rot, we
obtain the stationary solution
\eqn\spamr{
ds^2 = x^2\,dt^2 - 2\omega x^2\,dt\,dy + (\omega^2x^2 -
b^2x^{2\alpha})\,dy^2 - b^2x^{2\alpha}\,dx^2.
}
Invariant spacetime properties, such as the curvature scalar \ric\ or the
integrated geodesic equation \geospa, being unaffected by coordinate
transformations, the Penrose diagrams for the stationary spacetimes
\spamr\ are the same as for the corresponding static spacetime \spame.
The only new feature induced by the transformation \rot\ is the appearance
of closed timelike curves (CTCs). The
circles $t =$ const., $x =$ const.\ are CTCs for $x > x_0 \equiv
(\omega/b)^{1/(\alpha-1)}$ if $\alpha < 1$, and for $x < x_0$ if $\alpha >1$.
Accordingly, CTCs occur in the region extending from the circle $\rho =
\rho_0 \equiv bx_0^{\alpha+2}/|\alpha+2|$ to the singularity $\rho = 0$
if $\alpha \leq -2$ or $\alpha > 1$, and between the circle $\rho =
\rho_0$ and spacelike infinity $\rho \rightarrow \infty$ for $-2 < \alpha
< 1$. For $\alpha = 1$, all the circles $t =$ const., $x =$ const.\ are
CTCs if $|\omega| > b$, and CLCs (closed lightlike curves) if $|\omega| =
b$. The metrics \spamr\ with $\alpha = 1$, $|\omega| < b$ do not admit
CTCs, and all describe the same static spacetime, as may be shown by
performing on \spamr\ the global coordinate transformation
\eqn\glob{
t = ( 1 - {{\omega^2}\over{b^2}} )^{1/2}\hat{t}, \quad x = b^{-1} 
\hat{x}, \quad y = b^{-1} ( 1 - {{\omega^2}\over{b^2}} )^{-1/2}
( \hat{y} - {{\omega}\over{b}}\hat{t} ) 
}
to the static conformally flat metric
\eqn\conf{
ds^2 = {{\hat{x}^2}\over{b^2}}(d\hat{t}^2 - d\hat{x}^2 - d\hat{y}^2).
}

\par
Now we consider the causal structure of the second static solution, Eq.\
\bh, which can be put in the Schwarzschild--like form  
\eqn\bhs{
ds^2 = f\,dt^2 - f^{-1}\,d\rho^2 - ( {{\k n^2\nu^2}\over{2}} )^2
\rho^2\,dy^2
}
with 
\eqn\fbh{
f = x^2 = {{2}\over{\k n^2\nu^2}}(B - \ln\rho^2).
}
The constant $B \equiv 2\ln(2b/\k n^2\nu^2)$ can be identified as a 
mass parameter. The global causal structure of this spacetime, considered
as a solution of the Einstein--Maxwell equations in (2+1) dimensions, has
previously been discussed in the case $\k > 0$ by Gott et al.\ \gsa, and
for both signs of $\k$ by Kogan \kogan. The Ricci tensor has a single
nonvanishing component
\eqn\scain{
R^2_2 = -{4\ov{\k n^2\nu^2\rho^2}},
}
showing that $\rho=0$ is the location of a curvature singularity, while
the Killing horizon corresponds to $\rho = \rho_h \equiv {\rm e}^{B/2}$.
The temperature associated with this horizon is
$\exp(-B/2)/(\pi|\k|n^2\nu^2)$ \reznik.  
Let us first consider the case $\k > 0$.
In the region $0<\rho<\rho_h$, where $t$ is a timelike coordinate,
there is a timelike  singularity at $\rho=0$ where the spacetime is
null, timelike and spacelike incomplete. In the region $\rho_h<\rho<\infty$,
on the other hand, $\rho$ becomes a timelike coordinate and the boundary 
$\rho=\infty$ is geodesically complete.
Following the simple rules given in \klosch\ we can construct the maximally 
extended Penrose diagram, which is represented in Fig.\ 7. 
As also suggested in \gsa\ the singularity structure of this solution
is very similar to that of the Reissner-Nordstr\"om solution and 
corresponds to two point charges with opposite values of the electric
charge. Also, $\rho=\rho_h$ does not correspond to an event horizon, but
to a Cauchy horizon similar to the inner horizon of the Reissner-Nordstr\"om 
solution.

\par
We find more interesting to interpret physically
the solution given by $\k<0$ (as first indicated in \kogan).
The analysis is similar to that of the previous case, except
for the important fact that the signature of the metric is changed:
$\rho=0$ is a spacelike singularity, $\rho=\rho_h$ is an event horizon
and $\rho=\infty$ is still infinitely distant. The Penrose diagram 
is identical to that of the Schwarzschild solution, see Fig.\ 8
(it is obtained by rotating the diagram of Fig.\ 7 by 90 degrees),
and therefore this solution represents a black hole.
This is probably the black hole solution of three--dimensional 
gravity which, at least for what concerns the causal structure,
is closest to its four--dimensional version, i.e. the Schwarzschild black hole.
In this case, the integrated geodesic equation reads
\eqn\mogeo{
\dot\rho^2 + V(\rho) =  E^2, \ \ \ V(\rho)= -{{2}\ov{\k n^2\nu^2}}
( \ln\rho^2 - B ) ( \varepsilon + {{\lambda^2}\ov{\rho^2}})
}
($\lambda = 2L/\k n^2\nu^2$). The form of $V(\rho)$ depends on the value of 
$\lambda$. For $|\lambda| < {\rm e}^{1+B/2}$, see Fig.\ 9, all timelike 
geodesics ($\varepsilon = +1$) get captured by the
black hole, i.e. their worldlines start at the past singularity
and end in the future singularity of Fig.\ 8. This peculiar behaviour,
not shared by all geodesics in the Schwarzschild spacetime,
is due to the fact that the static frame $(t, \rho)$ is not
asymptotically inertial, i. e. the black hole will not be seen at rest
relative to an inertial observer at infinity. A similar phenomenon 
has been shown to exist in \balfab\
for solutions to $1+1$ dimensional dilaton gravity representing
black holes which are static only as viewed by asymptotic
accelerated (Rindler) observers. 
In the case $|\lambda| > {\rm e}^{1+B/2}$, on the other hand, we show 
in Fig.\ 10 that, due to the `high' angular momentum, bounded motion, in 
particular also circular orbits, is possible.

\par
The stationary solution generated from the static solution \bhs\ by the
local transformation \rot\ is
\eqn\bhr{
ds^2 = f\,dt^2 - 2\omega f\,dt\,dy + (\omega^2f - ({{\k
n^2\nu^2}\ov{2}})^2\rho^2)\,dy^2 - f^{-1}\,d\rho^2.
}
Again, the global causal structure of these stationary spacetimes is the
same as that of the original static spacetime, except for the appearance
of CTCs in the regions where $g_{yy}$ may become positive. For $\k > 0$, 
$g_{yy}$ always has a zero at some $\rho = \rho_1 < \rho_h$, and CTCs occur 
between
the circle $\rho = \rho_1$ and the singularity at $\rho = 0$. For $\k < 0$,
$g_{yy}$ has a maximum at $\rho = \rho_m \equiv \omega(-\k n^2\nu^2/2)^{-3/2}$;
if $\rho_m < {\rm e}^{(1+B)/2}$, $g_{yy}$ is
negative at $\rho = \rho_m$ and so everywhere outside the horizon, and
there are no CTCs; if $\rho_m > {\rm e}^{(1+B)/2}$, $g_{yy}$ vanishes on
two lightlike circles outside the horizon, and CTCs occur in the region
between these two circles.

\newsec{Multibody structure}

The general static multicenter solution to Eqs. \compa\--\erpa\  can 
in principle be constructed by replacing in \hh\ the
one--particle solution $w = \zeta \equiv x + iy$ (where $x$ and $y$ are the
radial and angular coordinate respectively) by
\eqn\mulce{
w = A_0 + \sum_{i=1}^N A_i\ln (z-a_i)
}
(where now $z = {\rm e}^{\zeta}$) for real weights $A_0$, $A_i$ and 
complex center locations $a_i$. The corresponding potential $\s$ solving
Eq. \sig\ is then (up to an additive constant)
\eqn\sigu{
\s_{(1)} = a\ln{\rm Re}w(z)
}
for the first class of solutions, and
\eqn\sigd{
\s_{(2)} = c\,{\rm Im}w(z)
}
for the second class. In this last case, the resulting $\s$--model field
$\psi(z)$ is single--valued only if all the $cA_i$ are integers,
\eqn\Ai{
A_i = {{n_i}\ov{c}}.
}
The metric function $u$ in \stato\ is then obtained by integrating Eq.
\inde, which may be written as 
\eqn\inda{
\dd_{z}u = {{1}\ov{2}}\dd_{z}\ln(\dd_{z}w) + 
\kappa\nu^2h(\dd_w\s)^2\dd_{z}w,
}
leading to the two classes of solutions
\eqn\multu{
ds_{(1)}^2 = h^2 dt^2 - b^2h^{\kappa\nu^2a^2}|w'(z)|^2dz dz^{*}
}
(where $h = {\rm Re}w(z)$, and $b$ is a constant), and
\eqn\multd{
ds_{(2)}^2 = h^2 dt^2 - b^2{\rm e}^{-\kappa c^2\nu^2h^2/2}|w'(z)|^2dz 
dz^* .
}

\par
This solution describes a distribution of $p$ black holes (where $p \le N$
is the number of disconnected components of the horizon $h = 0$) under the
same conditions as for the one--particle solutions, i.e. $\kappa\nu^2a^2 =
4(1-n)/n$ ($n$ positive integer) for the first class of solutions, and
$\kappa < 0$ for the black holes of the second class. However, 
besides the $p$ black holes, additional
conical singularities will in general be present at the $(N-1)$
zeroes $z_j$ of the function $w'(z)$ \dj\ \mbh. 
It is in principle possible to choose the parameters in \mulce\
so that these conical singularities are absent, which is the case if the
parameters are constrained by the $2(N-1)$ relations
\eqn\conabs{
\sum_{i=1}^nA_ia_i^q = 0,
}
for $q = 1, \cdots, N-2$ (i.e. all the multipole moments until the $2^{N-2}$ 
order vanish). But the function $w'(z)$ will still have a pole of
order $N$, leading for $N > 2$ to a conical singularity at $z \rightarrow 
\infty$ 
(but at finite geodesic distance) of the metric \multu\ or \multd. In the 
case $N = 2$ the regular solution is 
\eqn\efdu{
w(z) = A_0 + A_1\ln\left( {{z-a_1}\ov{z-a_2}} \right);
}
the corresponding horizon 
\eqn\hord{
{{|z-a_1|}\ov{|z-a_2|}} = {\rm e}^{-A_0/A_1}
}
being a circle, the resulting one--black hole spacetimes are identical
to those of Sect.\ 3.

\par
So for $N > 2$ the solutions \multu\ or \multd\ always admit conical
singularities. As conical singularities correspond in $2+1$ gravity
to point particles we will require, for consistency, that
their worldlines are geodesics of the spacetime \dj. 
Such freely falling particles momentarily at rest  ($v^i
\equiv dx^i/dt = 0$) can remain at rest only if the Newtonian force
\eqn\stage{
{{dv^i}\ov{dt}} = {{1}\ov{2}}g^{ij}\dd_jg_{00}= -{\rm e}^{-2u}{h\dd_i h}
}
vanishes at their location. In the case of the
above multi--center solutions \multu\ \multd, this is possible only if
the conical singularities lie on the horizon, i.e. if the parameters in
\mulce\ are constrained by the ($N-1$) relations
\eqn\conhor{
h(z_j) = 0.
}
The horizon world--sheet being generated by null geodesics, it follows
that the conical singularities at
$z = z_j$ do lie on null geodesics when the conditions \conhor\ are satisfied.
Under these conditions, the generic $N$--center solution, which has a
single horizon with ($N-1$) self--intersections, is seen by an
outside observer as a static system of $N$ black holes and ($N-1$) conical
singularities.

\par
The global structure of such multi--black hole spacetimes depends on the
analytical extensions which are performed. Let us discuss for definiteness
the case $N=2$, $A_1=A_2$ (symmetrical two--black hole). In this case the
horizon makes a figure 8. To each half of this 8 we can glue a distinct
``interior'' region II of the extended one--black hole spacetime. Then, we
can glue the other horizon of each of these regions II to one of the two
halves of the figure 8 horizon in an exterior two--black hole region
identical to the first one. Such a symmetrical extension can easily be
generalized to a $N$--black hole spacetime made of two identical exterior
regions connected by $N$ Einstein--Rosen--like bridges \er. In a more
economical extension ($N=2$), similar to the Wheeler--Misner construction of
\flat, the two exterior regions are identified, i.e. the two horizons
bounding a single interior region II are glued to the two horizons of the
same exterior region; this construction can easily be generalized to any
even $N$, with a single interior region and $N/2$ interior regions II.  
In the case of the first class of black hole solutions with $n$ even,
where the one--black hole regions II are isometrical to the regions I, the
simplest causal symmetrical extension is achieved, for any $N$, by gluing
the $N$ future horizons of an exterior region to the $N$ past horizons of
the next exterior region. The resulting generalized Penrose diagram is
similar to the diagram (Fig.\ 6) for the one--black hole spacetime, with
multiple lines at 45$^o$ standing for the multiple horizon components \mbh.

\par
While consistent, such static multi--black hole solutions seem rather
special, as one would intuitively expect that black holes should attract (and
therefore fall on) each other. However, following \mbh\ one can generalize
these static solutions to consistent dynamical solutions of the Einstein--$\s$
equations by taking the N complex parameters $a_i$ in \mulce\ to be
time--dependent, 
\eqn\time{
a_i = a_i(t),
}
and requiring that the $(N-1)$ conical singularities $\zeta_j(t)$ follow
geodesics of the resulting spacetime. The unknown functions $a_i(t)$ are
then determined from given initial conditions, up to an arbitrariness
corresponding to that of the center--of--mass motion of the system.

\par
Let us show for definiteness how to construct such a dynamical solution
corresponding to a system of two black holes with equal masses. Choose a
particular geodesic $w = w_1(t)$ in a one--black hole spacetime $w =
\zeta$ of Sect.\ 3 , and make the global coordinate transformation
\eqn\transf{
z^2 = c[{\rm e}^{w/\alpha} - {\rm e}^{w_1(t)/\alpha}]
}
($c$, $\alpha>0$ real constants). The time--dependent field configuration 
$(\s(z,t), g_{\mu\nu}(z,t))$ transformed from the static $N=1$ solution
$(\s(w), g_{\mu\nu}(w))$ by the coordinate transformation \transf\ is, by
construction, a solution of the Einstein--$\s$ equations, with the conical
singularity $z = 0$ following a geodesic. This solution is of the form
\mulce\ with $N=2,\, A_0 = -\alpha\ln c,\, A_1 = A_2 = \alpha,\, a_1(t) =
-a_2(t) = \sqrt{-c}\,{\rm e}^{w_1(t)/2\alpha}$. The static solution
discussed above is recovered if we choose the special null geodesic
$\varepsilon = L = E =0$ in Eq.\ \geospa\ or \mogeo.

\par
We briefly discuss the dynamical evolution of such two--black hole systems
generated from the different types of one--black holes encountered in this
paper. In the case of the first--class black holes with $n$ odd ($n > 1$;
the case $n=1$ is treated in \mbh), the consideration of the generic
timelike or null geodesic\foot{The case where the conical singularity
follows a spacelike geodesic, $\varepsilon = -1$, would lead to tachyonic
two--black hole systems.} leads to the following picture. A 
distant observer sees a single past horizon suddenly developing a conical
singularity and bifurcating. The two horizons then separate to a finite
distance and merge again (fall back on each other) after an infinite
coordinate time --- but a finite proper time for the distant freely
falling observer, who theoretically should see the merger at the same
instant he or she crosses the resulting single future horizon into the
black hole (all timelike geodesics \geospa\ cross the horizon $\rho = 0$).
Actually our hypothetical three--dimensional observer will not live long
enough, having been stretched apart and destroyed by infinitely rising
tidal forces before being able to cross this horizon of infinite length
\cold.

\par
The historical picture is the same in the case of the first--class black
holes with $n$ even, $n = 2q$. However, in this case Re$(w) = x =
(\rho/b)^q$ stays real when the horizon is crossed ($\rho \to -\rho$), so
that one can, at least formally, analytically continue \transf\ across the
horizon. For $q$ even, Re$(w)$ does not depend on the sign of $\rho$, so
that the dynamical evolution is the same in the sectors $\rho > 0$ and
$\rho < 0$. On the other hand, for $q$ odd, Re$(w)$ changes sign with
$\rho$. For $\rho< 0$, the line at spacelike infinity $\rho \to -\infty$ of the
one--black hole solution is mapped by \transf\ into the two lines
$z_{\infty}(t) = \pm a_1(t)$. So in a sector $\rho < 0$ there are two regions
at spacelike infinity. A distant observer in one of these regions sees a
conical singularity suddenly appearing on his past horizon, which merges
with the past horizon of the other region at spacelike infinity. The
subsequent spacelike (or lightlike) sections of this universe are
``trousers'' with two legs connected (at the conical singularity) to one
trunk ending on a single horizon. Finally, both the conical singularity
and the observer (who is destroyed in the process) fall back on the
horizon. 

\par
The case of second--class black holes differs in several respects.
Observers can now cross the horizon (of finite length) unharmed.
Also, distant freely falling observers can avoid altogether falling
into the black hole if they have enough angular momentum. Furthermore,
there are now three possible dynamical evolutions for a two--black hole
system, according to the nature of the timelike or null geodesic followed
by the conical singularity. The first possible evolution is
similar to that described for the first--class systems, except that the
two horizons never actually merge for our distant observer in stationary
orbit. In the second scenario, corresponding to a class of null geodesics,
the two black holes, infinitely separated at $t = -\infty$, fall upon each
other at the speed of light, eventually merging at $t = +\infty$. The
third possibility, corresponding to a bounded motion of the conical
singularity, is that of a stationary system of two black holes orbiting
around their common center of mass (the conical singularity)\foot{Such
stationary sytems of two black holes with a conical singularity also occur
in the case of extreme BTZ black holes \mbh.}.
   

\vfill\eject

\newsec{APPENDIX}

In this Appendix, we discuss briefly the extension of the various static
solutions (with $\k\nu^2= -1/2$) of the three--dimensional Einstein--$\s$
equations derived in Sect.\ 3 to ${\cal E} = p^2$ solutions of the
(3+1)--dimensional Einstein--Maxwell equations. 

\par
In the case $h = 1$, $\psi = \psi(\zeta)$ \worm, Eq.\ \umpa\ is solved by
\eqn\usol{
{\rm e}^{2u} = (1+|\psi|^2) |k(\zeta)|^2,
}
where $k(\zeta)$ is an arbitrary analytical function of $\zeta$. This
function can be absorbed into a redefinition of the complex variable
$\zeta = x + iy$, leading to the three--dimensional Euclidean metric
\eqn\tresol{
ds_{(3)}^2 = (1+|\psi|^2)(dx^2 + dy^2) + dz^2.  
}
The corresponding stationary four--dimensional metric solving the
Einstein--Maxwell equations with ${\cal E} = 1$ is, from \statu,
\eqn\quasol{
ds_{(4)}^2 = (1+|\psi|^2) (dt - \omega_3\,dz)^2 - (1 +
|\psi|^2)^{-1}\,ds_{(3)}^2, 
}
where the potential $\omega_3$ solves Eq.\ \twist\ which reduces, for
${\cal E} = 1$ and a static three--dimensional metric, to
\eqn\twistd{
\dd_{\zeta}\,\omega_3 = h^{-1}\,(1+|\psi|^2)^{-2}\,(\psi^*\dd_{\zeta}\psi -
\psi\dd_{\zeta}\psi^*), 
}
leading in the present case to
\eqn\soltw{
\omega_3 = -(1+|\psi|^2)^{-1}.
}
We thus arrive at the four--dimensional metric
\eqn\pp{
ds_{(4)}^2 = (1+|\psi|^2)\,dt^2 + 2\,dt\,dz - dx^2 - dy^2,
}
which corresponds to a subclass of $pp$--wave spacetimes \exact.

\par
In the other case treated in Sect.\ 3, $h = x$ and $\psi$ is assumed to
depend on a single real potential $\sigma$. As discussed at the end of
Sect.\ 2, the corresponding four--dimensional Einstein--Maxwell solution
is singular in the case of the meridian ansatz \meri. We will consider
here only the case of the equator ansatz $\psi = {\rm e}^{i\sigma}$ with
${\cal E} = 1/2$. The resulting four--dimensional metric is then \staxi\
with $f = 1$ and, from Eqs. \elma\ and \twistm,
\eqn\twistre{
\dd_i\omega_3 = - \rho \epsilon_{ij}\,\dd_j\sigma.
}

\par In the case of the three-dimensional metric \spame\ with $\sigma =
a\ln x$, we thus obtain the four--dimensional metric and  
electromagnetic potentials
\eqn\qug{
ds_{(4)}^2 = (dt - az\,d\varphi)^2 - {b^2}\rho^{-a^2/2}(\,d\rho^2 + dz^2) -
\rho^2\,d\varphi^2,
}
\eqn\que{
v = {1\ov{\sqrt{2}}}\cos(a\ln\rho), \qquad u = {1\ov{\sqrt{2}}}\sin(a\ln\rho).
}
The curvature invariant $R^{\mu\nu}R_{\mu\nu} \propto \rho^{a^2-4}$ is
singular at $\rho = 0$ if $|a| < 2$, and at $\rho \rightarrow \infty$ if
$|a| > 2$. The case $|a| = 2$, 
\eqn\regqug{
ds_{(4)}^2 = (dt - 2z\,d\varphi)^2 - {{b^2}\ov{\rho^2}}(\,d\rho^2 + dz^2) -
\rho^2\,d\varphi^2,
}
corresponds to the regular homogeneous solution of the Einstein--Maxwell
equations previously obtained by McLenaghan and Tariq \mcle\ and Tupper 
\tupper. The axis $\rho = 0$ is at infinite geodesic distance; let us also 
mention that all
the circles $t$, $\rho$, $z$ constant are timelike for $4z^2 - \rho^2 > 0$.

Similarly, the three--dimensional metric \bh\ with $\sigma = -2ay$ ($a =
-n/2$ real) leads to the four--dimensional metric and electromagnetic 
potentials
\eqn\qbg{
ds_{(4)}^2 = (dt - a\rho^2\,d\varphi)^2 - b^2{\rm e}^{a^2\rho^2}
(\,d\rho^2 + dz^2) - \rho^2\,d\varphi^2,
}
\eqn\qbe{
v = {1\ov{\sqrt{2}}}\cos(2az), \qquad u = -{1\ov{\sqrt{2}}}\sin(2az),
}
previously given by McIntosh \mcin. The metric \qbg\ admits CTCs for
$\rho > |a|^{-1}$. A common feature of the solutions \regqug\ and \qbg\ is
that in both cases the Maxwell field does not share the spacetime symmetry
\mcin\ \exact.

\vfill\eject
\listrefs
\vfill\eject

{
 \epsfxsize=5.4cm 
\epsfysize=7.65cm
 \epsfbox{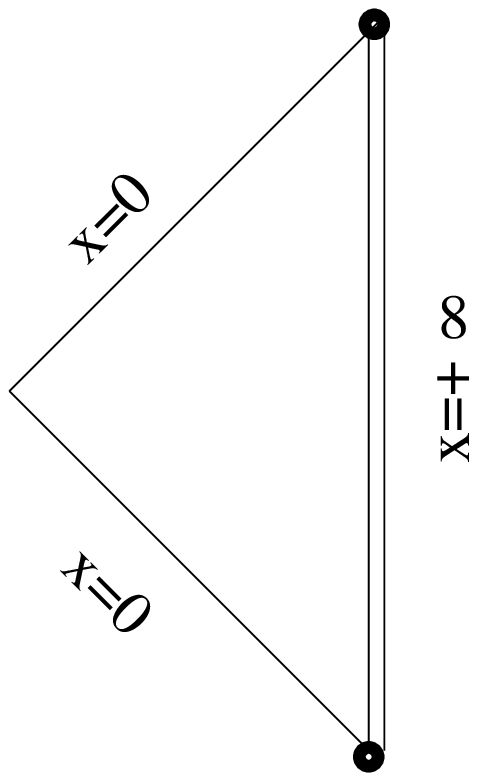}
 }
{\bf {Fig.\ 1:}}{
Penrose diagram for the spacetime (4.2) with $\alpha \le -2$. The radial
coordinate $x = ({g_{tt}}^{1/2}$ is related to the coordinate $\rho$ of
(4.2) by $\rho = bx^{\alpha+2}/|\alpha+2|$. The
singularity is represented by a double line.}

{
 \epsfxsize=5.4cm 
\epsfysize=7.65cm
 \epsfbox{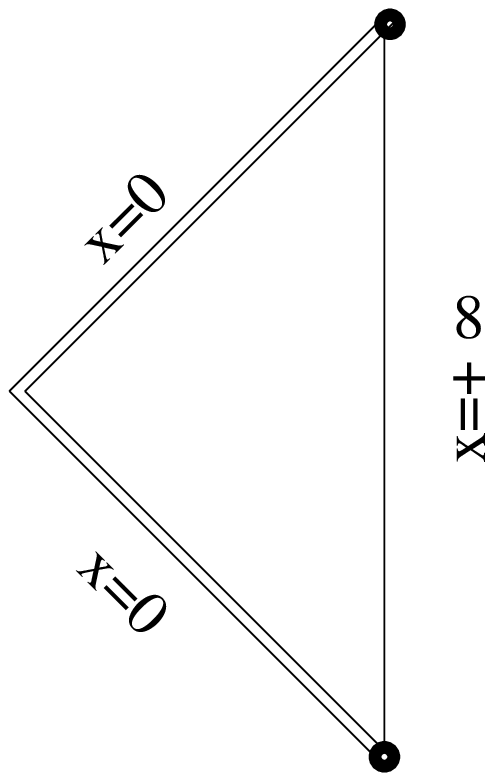}
 }
{\bf {Fig.\ 2:}}{
Penrose diagram for the spacetime (4.2) with $-2 < \alpha < 0$.}
\vfill\eject

{
 \epsfxsize=8cm 
\epsfysize=8.5cm
 \epsfbox{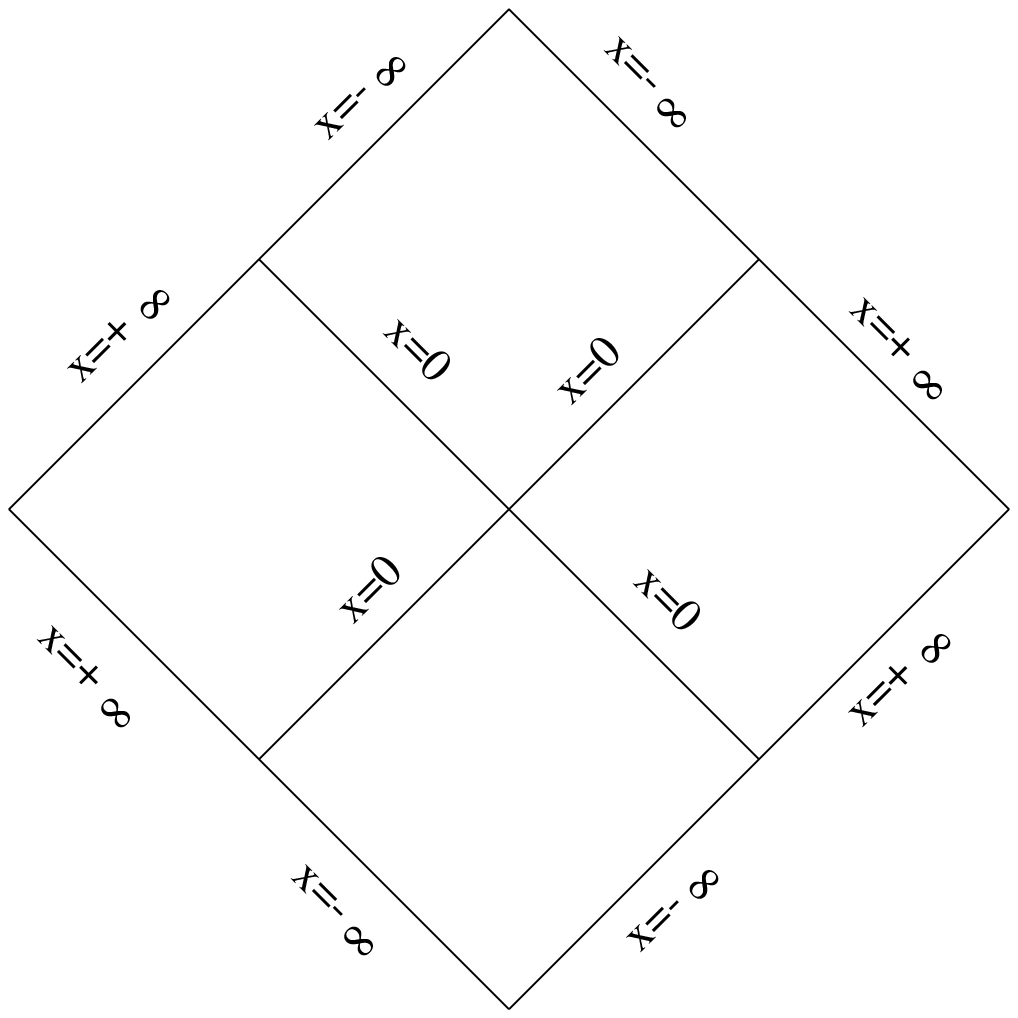}
 }
{\bf {Fig.\ 3:}}{
Penrose diagram for the spacetime (4.2) with $\alpha = 0$.}

{
 \epsfxsize=4.5cm 
\epsfysize=8.5cm
 \epsfbox{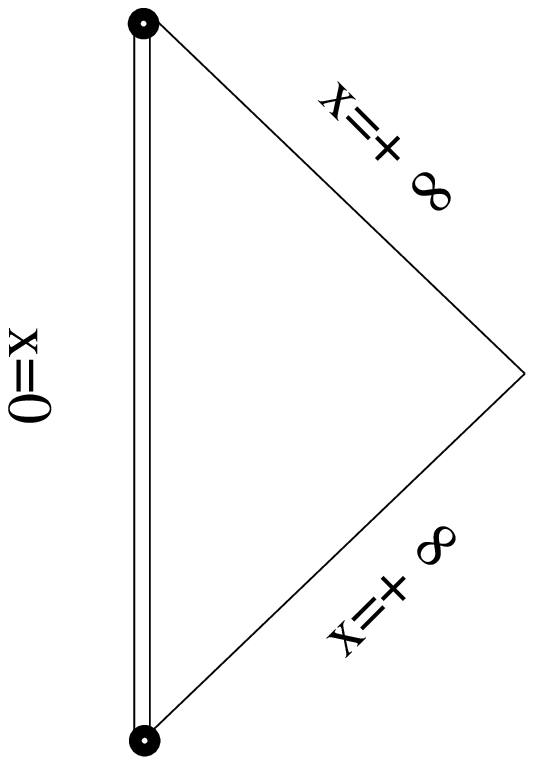}
 }
{\bf {Fig.\ 4:}}{
Penrose diagram for the spacetime (4.2) with $\alpha > 0$.}
\vfill\eject

{
 \epsfxsize=8cm 
\epsfysize=8.5cm
 \epsfbox{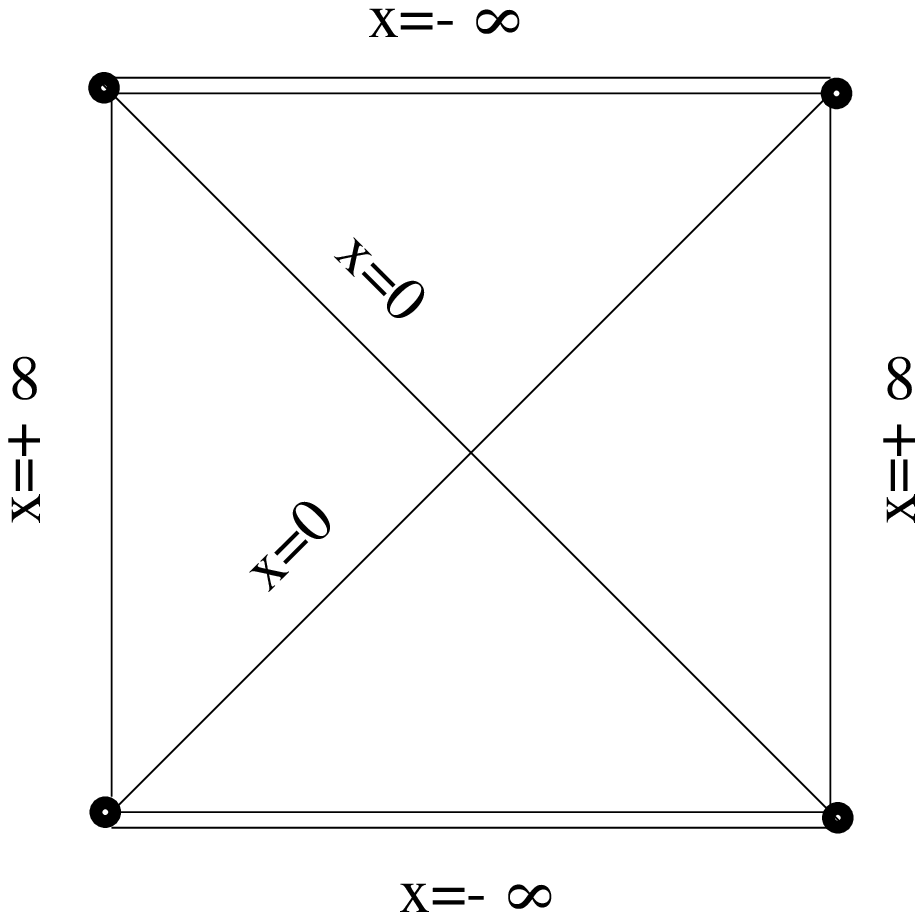}
 }
{\bf {Fig.\ 5:}}{
Penrose diagram for the first--class black hole with $n$ odd.}

{
 \epsfxsize=4cm 
\epsfysize=8.5cm
 \epsfbox{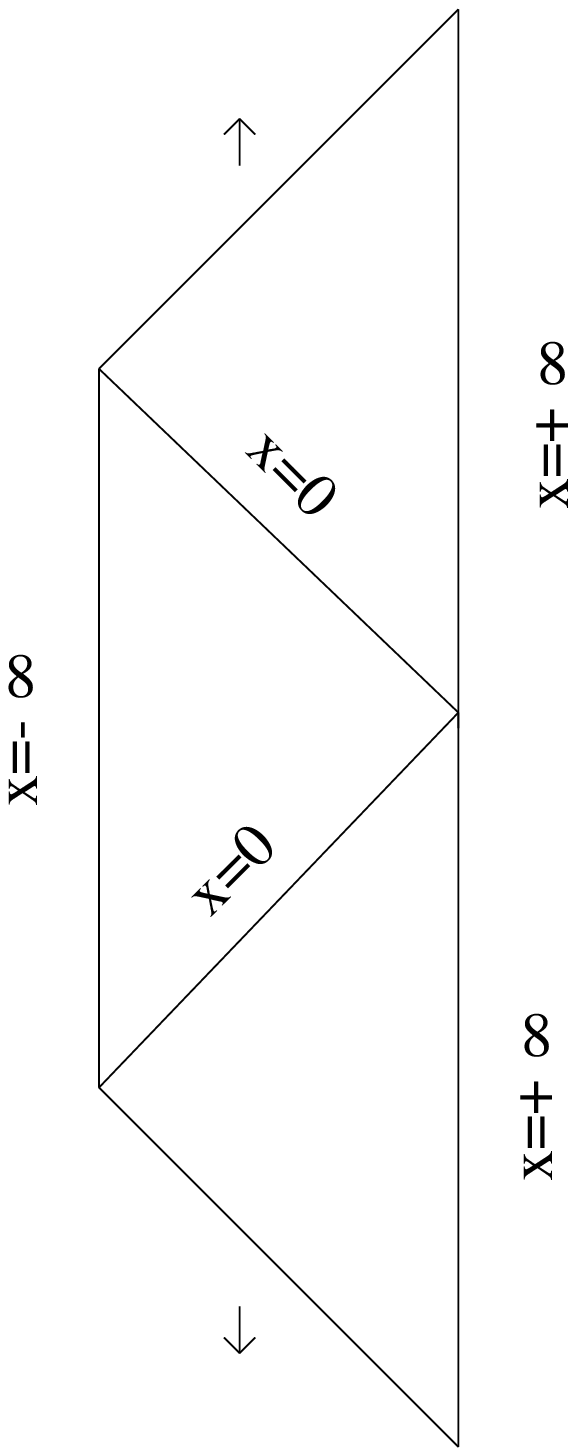}
 }
{\bf {Fig.\ 6:}}{
Penrose diagram for the first--class black hole with $n$ even.}
\vfill\eject

{
 \epsfxsize=7cm 
\epsfysize=8cm
 \epsfbox{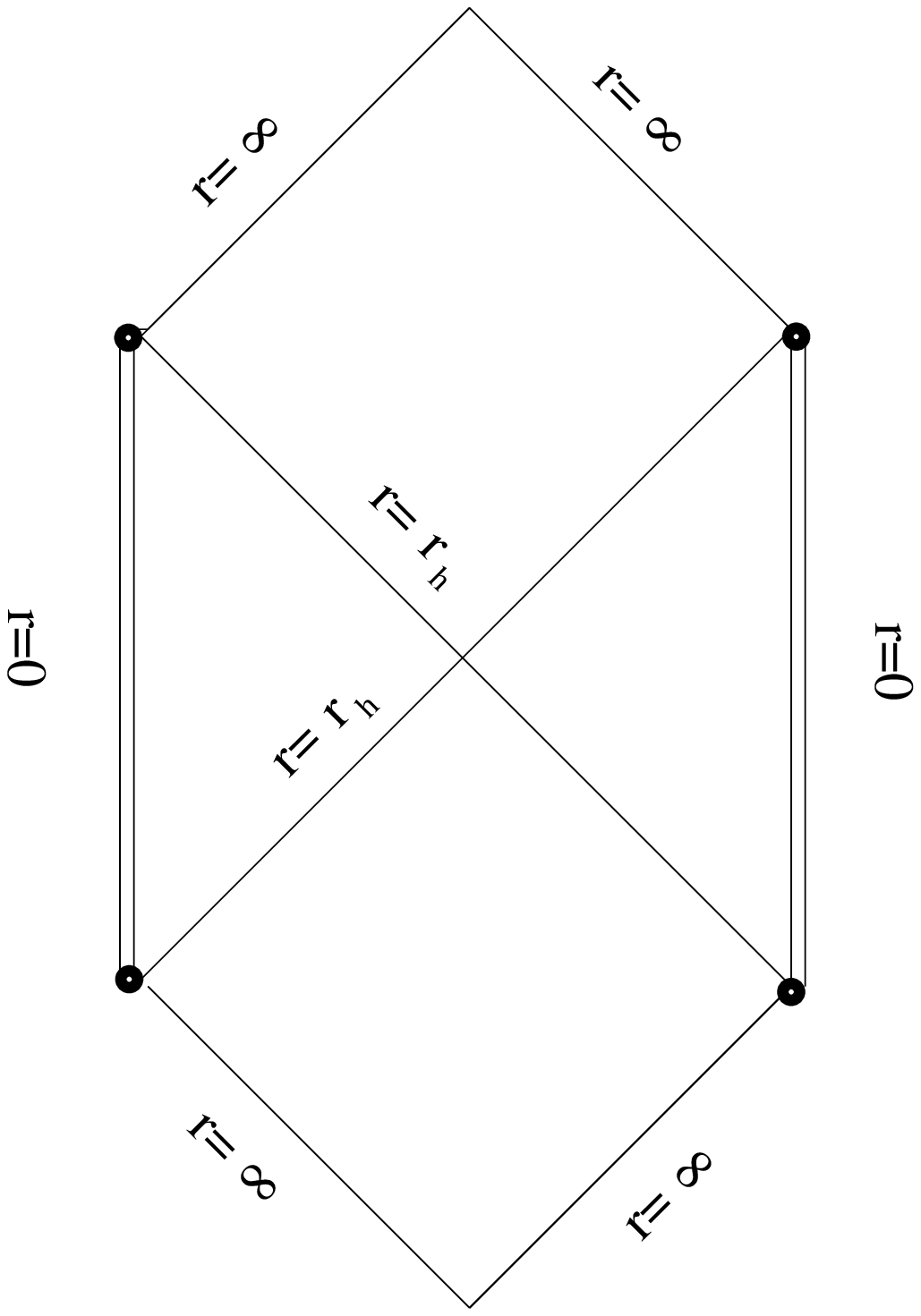}
 }
{\bf {Fig.\ 7:}}{
Penrose diagram for the spacetime (4.14)-(4.15) with $\kappa > 0$.}

{
 \epsfxsize=9cm 
\epsfysize=8.5cm
 \epsfbox{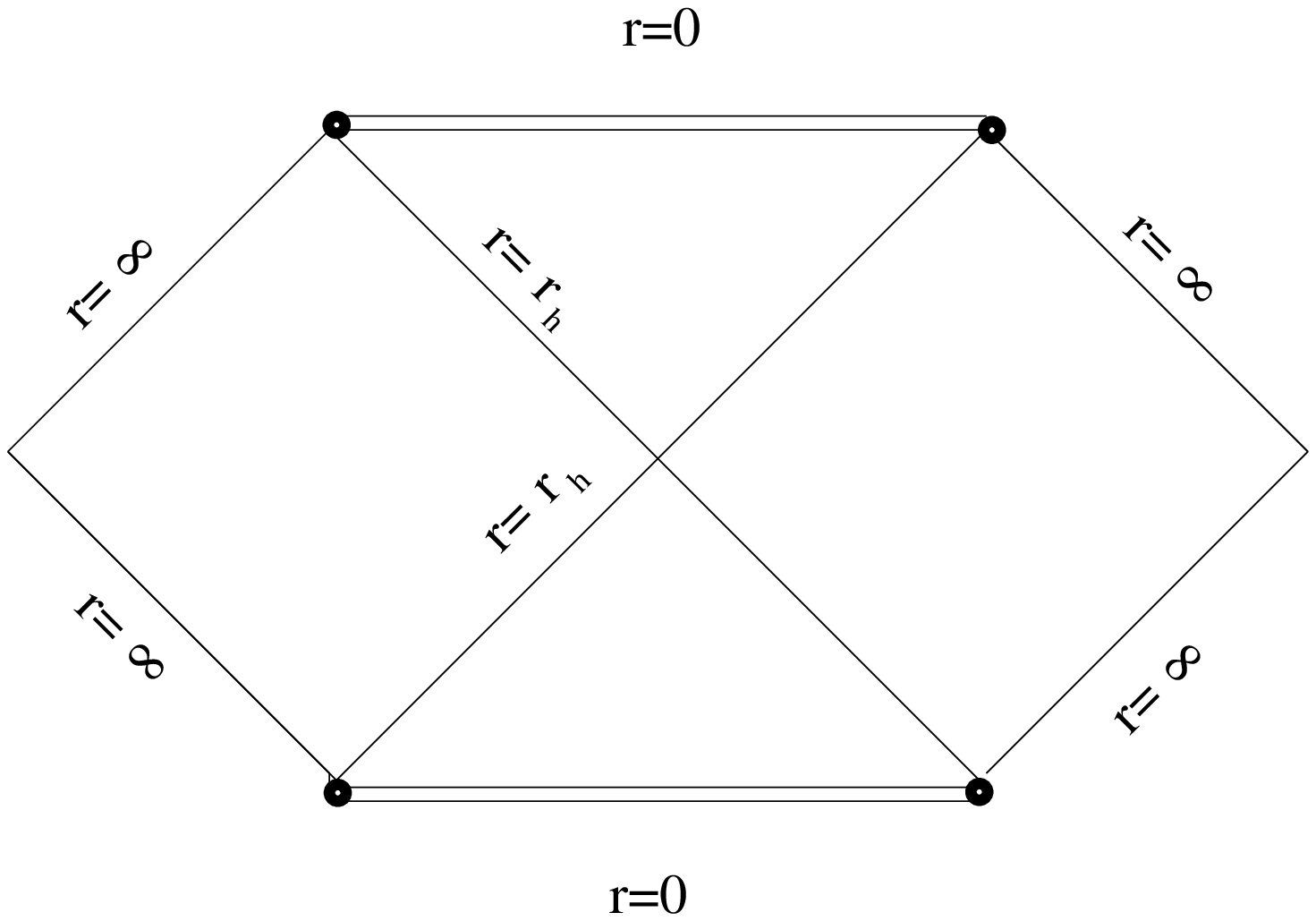}
 }
{\bf {Fig.\ 8:}}{
Penrose diagram for the second--class black hole (spacetime (4.14)-(4.15) 
with $\kappa < 0$).}
\vfill\eject

{
 \epsfxsize=12cm \epsfysize=16cm
 \epsfbox{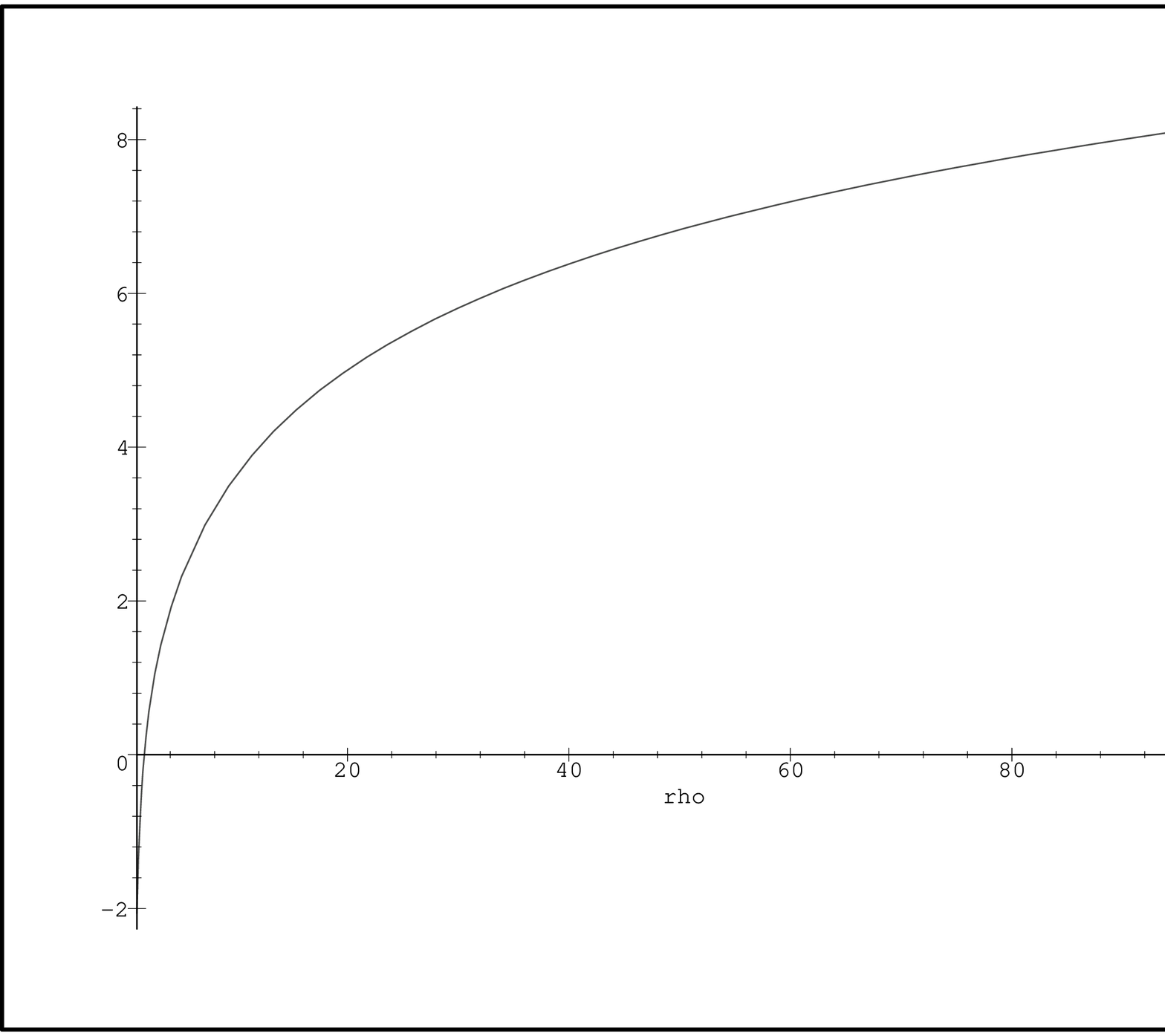}
 }
{\bf {Fig.\ 9:}}{
Graph of the potential $V(\rho)$ (Eq.\ (4.17)) for $\epsilon = +1$, $B=1$
and $\lambda = 1$.}
\vfill\eject

{
 \epsfxsize=12cm \epsfysize=16cm
 \epsfbox{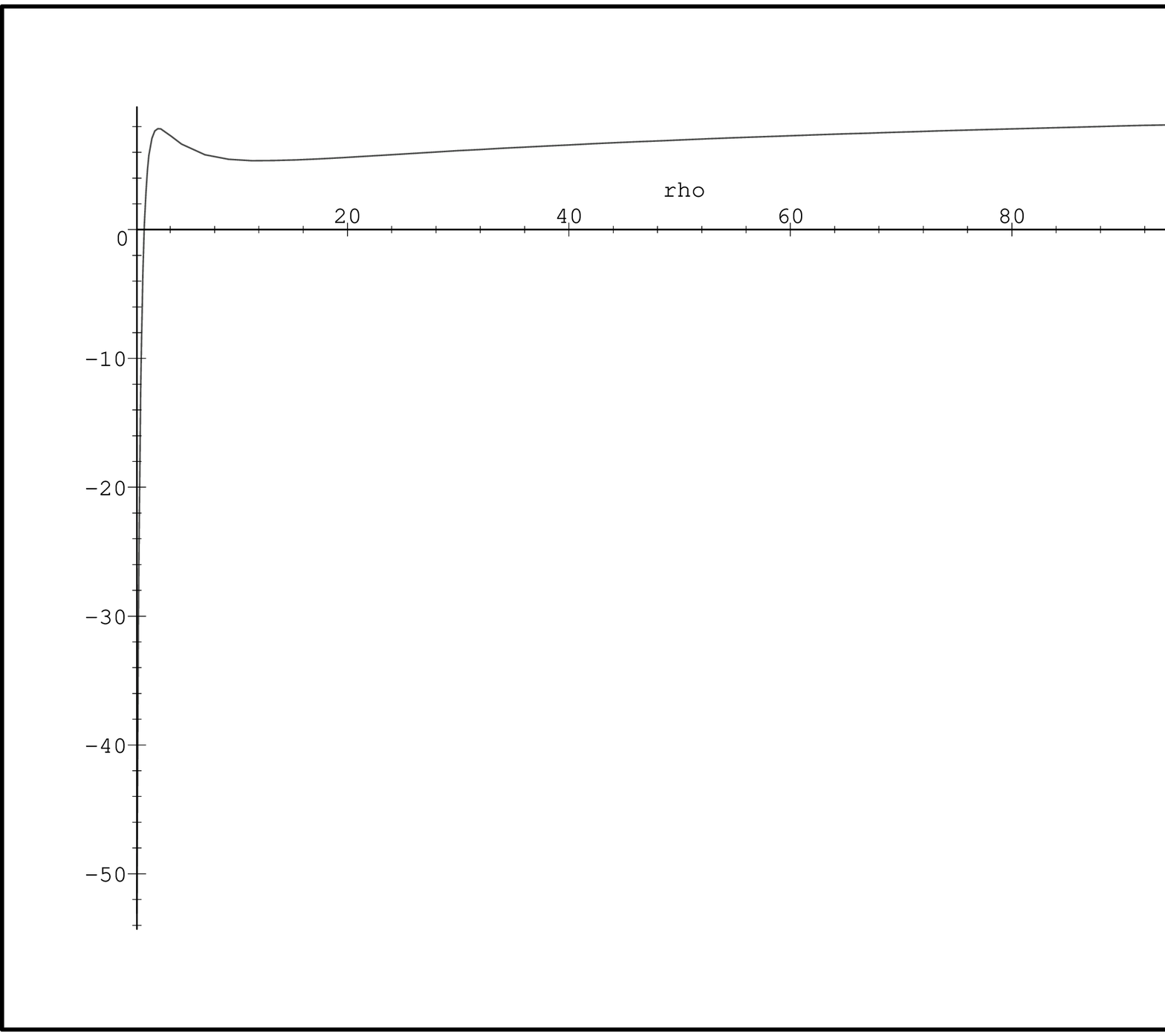}
 }
{\bf {Fig.\ 10:}}{
Graph of $V(\rho)$ for $\epsilon = +1$, $B=1$ and $\lambda = 50$ 
(case $\lambda > {\rm e}^{1+{B\ov 2}})$.}
\vfill\eject

\end